\documentclass[a4paper]{spie}  

\usepackage{amsmath,amsfonts,amssymb}
\usepackage{graphicx}
\usepackage{xspace}
\usepackage[colorlinks=true, allcolors=blue]{hyperref}
\usepackage[sort,super]{natbib}
\usepackage{eurosym}

\newcommand{\uas}{\ensuremath{\mu\mbox{as}}\xspace}

\newcommand{\msun}{\rm M_\odot} 

\newcommand{\gaia}{\textit{Gaia}\xspace}

\newcommand{\theia}{\textit{Theia}\xspace}

\newcommand{\hips}{\textit{Hipparcos~}}
\newcommand{\hst}{\textit{HST}}

\newcommand{\simMission}{\textit{SIM}}

\newcommand{\planck}{\textit{Planck}} 
 
\newcommand{\euclidx}{\textit {Euclid}\xspace} 
\newcommand{\euclid}{\textit {Euclid}} 
 
\newcommand{\herschelx}{\textit {Herschel}\xspace}
 
\newcommand{\jwst}{\textit{JWST}} 
\newcommand{\wfirst}{\textit{Roman}}
\newcommand{\rubin}{\textit{Rubin}} 
\newcommand{\ligo}{\textit{LIGO}}
\newcommand{\lisa}{\textit{LISA}}

\title{Theia : science cases  and mission profiles for high precision astrometry in the future}

\author[a]{Fabien Malbet}
\affil[a]{Univ.\ Grenoble Alpes, CNRS, IPAG, 38000 Grenoble, France}
\author[b]{Lucas Labadie}
\affil[b]{Univ.\ of  Cologne, Cologne, Germany}
\author[c]{Alessandro Sozzetti}
\affil[c]{Obs.\ Torino/INAF, Pino Torinese, Italy}
\author[d]{Gary A.\ Mamon}
\affil[d]{Sorbonne Université, CNRS, Institut d’Astrophysique de Paris, Paris, France}
\author[e]{Mike Shao}
\author[e]{Renaud Goullioud}
\affil[e]{Jet Propulsion Laboratory, California Institute of  Technology, Pasadena, CA, USA}
\author[f]{Alain Léger}
\affil[f]{Univ.\ Paris-Saclay, CNRS, Institut d'astrophysique spatiale, Orsay, France}
\author[c]{Mario Gai}
\author[c]{Alberto Riva}
\author[c]{Deborah Busonero}
\author[i]{Thierry Lépine} \affil[i]{Institut d'Optique \& Hubert
  Curien Lab, Univ.\ de Lyon, Saint-Etienne, France}
\author[a]{Manon Lizzana}
\author[g]{Alexis Brandeker}
\affil[g]{Stockholm Univ.\,Stockholm, Sweden}
\author[h]{Eva Villaver} \affil[h]{Centro de Astrobiología (CAB, C
  SIC-INTA), ESAC Campus, s/n/, 28692 Villanueva de la Cañada, Madrid,
  Spain}

\authorinfo{Send correspondence to FM using the email address
  \texttt{Fabien.Malbet@univ-grenoble-alpes.fr}}

\begin{document} 
\maketitle

\begin{abstract}
  High-precision astrometry well beyond the capacities of Gaia will
  provide a unique way to achieve astrophysical breakthroughs, in
  particular on the nature of dark matter, and a complete survey of
  nearby habitable exoplanets. In this contribution, we present the
  scientific cases that require a flexibly-pointing instrument capable
  of high astrometric accuracy and we review the best mission profiles
  that can achieve such observations with the current space technology
  as well as within the boundary conditions defined by space
  agencies. We also describe the way the differential astrometric
  measurement is made using reference stars within the field. We show
  that the ultimate accuracy can be met without drastic constrains on
  the telescope stability.
\end{abstract}


\section{INTRODUCTION}
\label{sec:intro}  

ESA's \hips and \gaia global astrometry scanning missions have
revolutionized our view of the Solar Neighborhood and Milky Way, and
provided crucial, new foundations for many disciplines of
astronomy. The topic of high precision astrometry arose from the
\emph{Space Interfermetry Mission (SIM)} in the late 2000's in order
to detect Earth-like exoplanets but keeping the methodology to achieve
absolute astrometry. Following the dismissal of SIM by NASA in 2010, a
small team decided to propose another concept to address high
precision astrometry called \emph{Nearby Earth astrometry Telescope
  (NEAT)}. NEAT consisting of a single off-axis mirror sending the
light onto a detector either in a formation flight
configuration\cite{Malbet+12} was presented for the M3 call for
missions at ESA. Then the \theia concept with a single spacecraft
carrying a Korsch three-mirror anastigmatic (TMA) telescope, a single
focal plane and instrument metrology subsystems has been studied and
submitted without success to M4, M5 and M7 calls for medium missions at
ESA. We present here a summary of the current concept status for
\theia as it has been submitted for the phase 1 call for M7 call for
missions.

The paper will quickly review the Science case, then present the
mission and the management structure before exploring in more details
how the most accurate measurements can be achieved using reference stars.

\section{Science case for high precision relative astrometry}

Differential astrometry can push the boundaries of our knowledge
further, as with this approach it is possible to achieve a degree of
relative positional precision in small fields (typically
$\leq 1^\circ$) vastly exceeding the levels obtained by a 10-yr \gaia
mission. This opens the door to $a)$ the identification of amplitudes
of astrometric variability phenomena in bright stars and $b)$ to the
determination of relative proper motions for faint stars with a
precision entirely out of the reach of state-of-the-art absolute
astrometry.

Ultra-high precision relative astrometry is identified as one of the
themes for a possible M-size mission in the final report of the senior
committee of ESA's \textit{Voyage 2050} long-term scientific plan. The
\theia micro-arsecond astrometric observatory is poised to achieve
unprecedented relative astrometric precision in small
fields\footnote{None of the scientific aims listed below require
  absolute astrometry.} that will bring breakthroughs on some of the
most critical questions of cosmology, exoplanetary science, and
particle physics. The details on the science cases have been described
in the White Paper published\cite{Malbet+21} for the \textit{Voyage 2050} report.

\subsection{Dark Matter }

While the mass density of the Universe is dominated by dark matter, we
know little of its nature, especially because dark matter particles
have not yet been observed in large accelerators (LHC).  \theia has
been designed to test the standard picture of collisionless, cold dark
matter, in the following ways:
 \begin{itemize}
 \item {\bf Test the cold nature of DM particles}.  Cosmological
   simulations of cold DM indicate that galaxy halos are full of
   subhalos\cite{Moore+99}.  Extrapolating the observed trends of
   $M_{\rm dyn}/L$ vs. $L$ to lower mass, subhalos below $10^8\,\msun$
   should be very or completely dark.  N-body simulations indicate
   that subhalos passing through the disk affect the kinematics of the
   stars in the neighborhood out to several kpc, leaving waves that
   persist for up to several hundred
   Myr\cite{Feldmann&Spolyar15,Malbet+21}. The proper motions measured
   by \theia at a dozen locations, located just above and below the
   galactic disk, should allow the first detection of kinematic
   imprints for several \emph{dark} subhalos\cite{Malbet+21}, which
   would be a major breakthrough.

 \item {\bf Test the collisionless nature of Dark Matter.}
   Cosmological simulations of DM particles indicate that structures
   (halos) have steep (``cuspy'') inner density
   profiles\cite{Navarro+96}. This has never been verified in
   galaxies, whose intermittent outflows from supernovae explosions
   and active galactic nuclei shake the inner gravitational potential
   (where the gas dominates), causing violent relaxation that rapidly
   leads the DM particles to settle to a more homogeneous inner
   density profile\cite{Pontzen&Governato12}. Dwarf spheroidal
   galaxies (dSphs) around the Milky Way (except the three most
   massive) are highly dominated by DM and should thus have cuspy
   density profiles. Analysis of proper motions adds two dimensions to
   the phase space probed by only using redshifts. Tests on mock data
   shows that this dramatically improves the measure of the inner
   slope of the density profile of dSphs\cite{Read+21,Malbet+21}. If
   \theia found cuspy inner density profiles in the targeted dSphs,
   this would confirm that DM is collisionless.  Conversely, if \theia
   found cores in the dSphs, this would provide the first measurement
   on the cross-section of self-interaction, a major breakthrough in
   particle physics.

 \item {\bf Determine the shape of the outer halo of the Milky
     Way}. Cosmological simulations of cold DM
   indicate\cite{Jing&Suto02} that galaxy halos have prolate outer
   shapes (while the stellar disks are oblate).  \theia will measure
   the proper motions of several distant known hyper-velocity stars
   thought to have received huge velocity kicks in three-body
   interactions involving the supermassive black hole at the center of
   the Milky Way\cite{Hills88}. These measurements combined with
   redshifts should recover the 3D directions of the motion\cite{Malbet+21} allowing
   to determine the halo axis ratios to 5\%, which is
   beyond the scope of \gaia.
\end{itemize}

These breakthroughs on dark matter will open up new directions in
Astrophysics, thus helping us understand the origin and composition of
the Universe.

\subsection{Early Universe}
\theia's precise astrometry will allow to probe the primordial
Universe in complementary ways.
\begin{itemize}
\item {\bf Test inflationary models and primordial black holes.} The
  power spectrum of primordial density fluctuations, which carries
  imprints from the initial inflation phase, is now well known on
  intermediate to large scales, thanks to the ESA's \planck\ cosmic
  microwave background mission, as well as large-scale surveys of
  galaxies. But it is poorly known at small scales (below 2 Mpc).
  Very high peaks from non-gaussian density fluctuations produced at
  phase transitions (e.g., QCD) or by features in the inflation
  potential collapse into primordial black holes\cite{Hawking71}
  (PBHs), while other high peaks can produce ultra-compact
  mini-halos\cite{Ricotti&Gould09} (UCMHs).  PBHs have been proposed
  as an alternative to an unknown particle to explain dark
  matter\cite{Frampton+10}. They could have masses down to
  $10^{-11}\,\msun$ (otherwise they would have evaporated by Hawking
  radiation\cite{Hawking71}).  The discovery of PBHs and/or UCMHs
  would constitute a major breakthrough, both for the existence of
  these objects and for constraints on inflationary
  models. Conversely, the absence of UCMHs would establish upper
  bounds on the amplitude of the primordial power spectrum on small
  scales\cite{Bringmann+12}.
   
  PBHs and UCMHs, if ubiquitous, could be detected by \theia by
  astrometric microlensing in the $\sim 20$ deep, regularly cadenced
  \theia fields ($5\,\rm deg^2$ in total).  Other astronomical objects
  may mimic as PBHs or UCMHs, but stars and brown dwarfs should be
  directly observable by launch date out to 1 kpc, drastically
  reducing the false detections.
    
\item {\bf Stochastic background of gravitational waves}.  The
  distortions to spacetime from gravitational waves (GWs) cause
  coherent distortions in the motions of stars in our Galaxy at
  locations that vary in time at the speed of
  light\cite{Moore+17} (0.3\,pc per year). The QCD transition
  produces a stochastic background of GWs whose low frequencies are
  inaccessible to \lisa\ or successors of \ligo, but well matched to
  \theia\cite{GarciaBellido+21} and \gaia\cite{Moore+17}.  \theia
  will be complementary to \gaia in searching for coherent proper
  motions: \gaia will have 700 times more sky coverage than \theia at
  galactic latitudes $|b| < 5^\circ$, but the detectability of
  coherent motions scales as proper motion precision over the square
  root of number of stars, and the 23 times higher proper motion
  precision of \theia's Deep Fields, will make it competitive with
  \gaia 10-year. Both \theia and \gaia should be more competitive than
  planned pulsar timing timing arrays (EPTA, etc.).
\end{itemize}

\subsection{Black holes and neutron stars}
\label{sec:nsbh}

While the detection by \ligo\ of gravitation waves caused by the
mergers of stellar-mass black holes (BHs) with other BHs or neutron
stars has opened an important new window on this class of compact
objects, there is much to be learned on the structure of neutron stars
and the many roles of BHs. \theia will strongly enhance our
understanding of compact objects in several ways.
\begin{itemize}
\item {\bf Merging supermassive black holes. } Merging of the most
  massive supermassive black holes also produce low-frequency GWs
  inaccessible by \lisa. While these events are rare, they are the key
  signatures of the build up of the most massive galaxies and their
  active galactic nuclei, and understanding whether the presence of
  supermassive black holes at very high redshifts ($z\sim8$) is caused
  by massive PBHs or by the buildup of non-PBH black holes. Just as
  for the case of the stochastic GW background, \theia should be
  complementary to \gaia for detecting merging supermassive BHs with a
  similar overall sensitivity, which should be superior to that of
  pulsar timing arrays.

\item {\bf Intermediate-mass black holes.}  Intermediate-mass black
  holes (IMBHs) are the fundamental link between stellar-mass BHs and
  supermassive ones lying in the cores of massive galaxies. But the
  kinematic evidence for IMBHs is sparse and debated, even in nearby
  globular clusters, with \hst\ and \gaia, because candidate IMBHs may
  instead be small, dark subclusters of compact stars or stellar-mass
  BHs\cite{Vitral+22}.  \theia's unprecedently precise proper motions
  (and parallaxes) will allow for many more central stars included in the
  mass modeling, which will allow a much more reliable distinction
  between IMBHs and dark subclusters. Observing with \theia at
  intermediate depth (200 total hours each) the half-dozen globular
  clusters with the greatest ratios of velocity dispersion to distance
  will provide the long awaited statistical view of the inner cores of
  globular clusters as a function of their mass, orbit around the
  Milky Way and core-collapse status.

\item {\bf Black Hole Formation \& Demographics.}  Astrometry offers a
  unique window into the BH population.  \theia could identify BHs by
  astrometric microlensing in the 20 deep fields aimed for other
  targets. \theia will also dedicate $\sim 15\%$ of its total
  observing time to follow up on alerts. These could be BH candidates
  from \wfirst, because \theia's higher temporal sampling compared to
  \wfirst\ is required to make precision mass estimates, through the
  combination of the magnification time series and the astrometry.
  The same data will provide excellent distance and space velocity
  information on the BHs.

  In X-ray binaries (XRBs), systematics in estimating binary
  inclination angles and distances limit the quality of BH mass and
  spin estimates. \theia will measure the wobble of inclination angles
  as well as the parallaxes for $\sim 50$ XRBs (most often too distant
  for \gaia).  These wobble measurements also yield the position
  angles of the orbits.  This will allow testing if the jet directions
  (from VLBI) are perpendicular to the binary orbital planes, and
  hence determine if the BHs were produced in bona fide supernovae
  with asymmetric mass loss, or by prompt collapse.

\item {\bf Constraining the equation of state of neutron stars.}
  Astrometric measurements of XRBs can yield a set of precise neutron
  star (NS) masses and constrain the NS equation of state (one of the
  core questions in nuclear physics) in three ways: (1) The maximum
  mass of a NS is set by the equation of state, meaning that finding
  heavier neutron stars eliminates certain equations of state. (2)
  X-ray pulse profile fitting is primarily sensitive to $M/R$ and
  distance, so that better masses and distances will provide more
  accurate radii. (3) The lowest mass NSs are likely formed from
  electron capture supernovae at the Chandrasekhar mass, thus allowing
  an estimate of the NS binding energy at the Chandrasekhar mass, and
  it is expected that these neutron stars should be in known classes
  of wind-fed XRBs.
\end{itemize}

\subsection{Earth-like planets around nearby sun-like stars}

The detection and atmospheric characterization of temperate,
potentially habitable telluric planets orbiting the nearest Sun-like
stars sits high amongst the key science themes both in the final
recommendations of the senior Scientific Committee for ESA’s long-term
scientific plan Voyage 2050 and in the long-term outlook of the
ASTRONET roadmap. \theia's surgical single-measurement positional
precision in pointed, differential astrometric mode ($< 1\,\mu$as)
will enable detection and high-confidence ($\geq 3\sigma$) true mass
determination for Earths and Super-Earths (M = 1--5 M$_\oplus$) in the
Habitable Zone (HZ) of the $\sim60$ nearest solar-type stars, using
high-cadence observations ($\approx 100$ visits) of each target and
$\geq 3$ reference stars\cite{Malbet+21}.

Extreme-precision, sub-m/s Doppler techniques from the ground are
expected to provide a global census of temperate terrestrial planets
around nearby late-type M dwarfs within the next 10--15
years. However, for solar-type primaries the Doppler method might be
limited by stellar activity and ultimately miss any HZ Earth-mass
companions whose orbits are not close to edge-on. \theia's astrometric
sensitivity will allow reaching three key goals in exoplanetary
science:
\begin{enumerate}
\item \theia astrometry will allow determining {\bf the true mass
    function of temperate 1--5 M$_\oplus$ rocky planets around
    solar-type stars}, {\em which is today completely unknown};
\item by measuring the true masses and full three-dimensional
  architecture in multiple systems \theia will allow studying for the
  first time {\bf the full demographics of planetary systems in the
    presence of temperate telluric planets around the nearest Sun-like
    stars}, in high synergy with \gaia and Doppler surveys;
\item the temperate telluric planets detected by \theia will
  constitute {\bf the fundamental input target list for direct-imaging
    / spectroscopic missions aimed at searching for atmospheric
    biomarkers}.
\end{enumerate}

It will be key for such ambitious space observatories either in the
optical / near-infrared (e.g., NASA's proposed flagship missions
HabEx, LUVOIR) or in the thermal infrared (e.g., the LIFE concept for
an ESA's L-class mission) providing their target list and avoiding
their research phase (about 50\% of their mission time) so that they
can invest all their precious observing time performing spectroscopy
of the atmospheres knowing exactly where to look. Prior knowledge of
the true masses will also crucially help in interpreting any molecular
detections in their atmospheres.

Updated estimates of \theia sensitivity\cite{Meunier&Lagrange22}
indicate that the median detectable mass across the full HZ for the
\theia\ stellar sample is $\simeq 1.1$ M$_\oplus$. If present-day
extrapolations of the occurrence rate of true Earth-like
planets\cite{Bryson+21} at $37_{-21}^{+48}\%$ are accurate, we then
expect to detect between 9 and 57 such planets. Furthermore, the
number of HZ Earths per star can be larger than unity, since a typical
HZ can dynamically sustain more than one planet.

\subsection{Basic scientific requirements}

\theia is a single field, visible-wavelength ($400-900$ nm)
differential astrometry mission, meaning that the derived astrometric
parameters for the target stars in a field will have position,
parallax and proper motion relative to a local reference frame tied to
a global one. At the time of the \theia mission, the most accurate and
complete optical reference frame will be that of the \gaia catalog. By
using \gaia global astrometry parameters as priors, the astrometric
solution of all the stars observed by \theia will be automatically
tied to the \gaia frame, without the need of forcing physical priors
on sources such as quasars or remote giant stars.  To cover the
science cases described above, two main observing modes are
necessary:\textbf{}
\begin{itemize}
\item \textbf{Faint Star Mode} (FSM) required by the Dark Matter and
  Compact Objects science cases, with $5\,\uas$/yr end-of-mission
  precision on proper motions for up to $>10^4$ science targets and
  references down to $R\simeq20$, observable within $\sim0.5^\circ$
  fields. The FSM uses the majority (exact trade-off to be determined)
  of a nominal 4-yr mission;
\item \textbf{Bright Star Mode} (BSM) required by the Exoplanets
  science case, with $1\,\uas$ single-measurement precision in 1-hr
  integration for the science targets and up to $10^2$ bright
  references with $R \leq 13$, observable within $\sim0.5^\circ$
  fields. The BSM utilizes the remainder of the observing time a
  nominal 4-yr mission not devoted to the FSM.
\end{itemize}

These characteristics make \theia much superior to all the
competition. The deep fields will achieve 23 times the proper motion
precision of \gaia (10
years)\footnote{https://www.cosmos.esa.int/web/gaia/science-performance\#astrometric\%20performance}
and 14 times that of \hst\cite{Vitral+22} (10 years). In the
bright-star regime\cite{Malbet+21}, \theia's 1-$\mu$as precision in
1-hr integration at the reference value $R=10$ mag exceeds that of
\gaia\cite{Lindegren+2018}, \wfirst\cite{sanderson+2019}, and {\em
  VLTI/GRAVITY}\cite{Gravity+2021} by factors of $\sim40-50$,
$\sim10-20$, and $\sim30-100$ respectively, and other
missions/instruments (e.g., \hst, \jwst, \emph{ELT/MICADO}, \rubin) by
even larger factors.

\section{Mission Configuration}
\label{sec:mission}

We describe quickly here the requirements, mission profile and
spacecraft design, resources and communications. We also assess the
technology readiness level (TRL) of the different components, the
development plans, fall-back scenarios, impact on science.  

\subsection{Payload and Platform Requirements}
\label{sec:prop-miss-prof}

\begin{figure}[t]
  \centering
  \includegraphics[width=0.99\hsize]{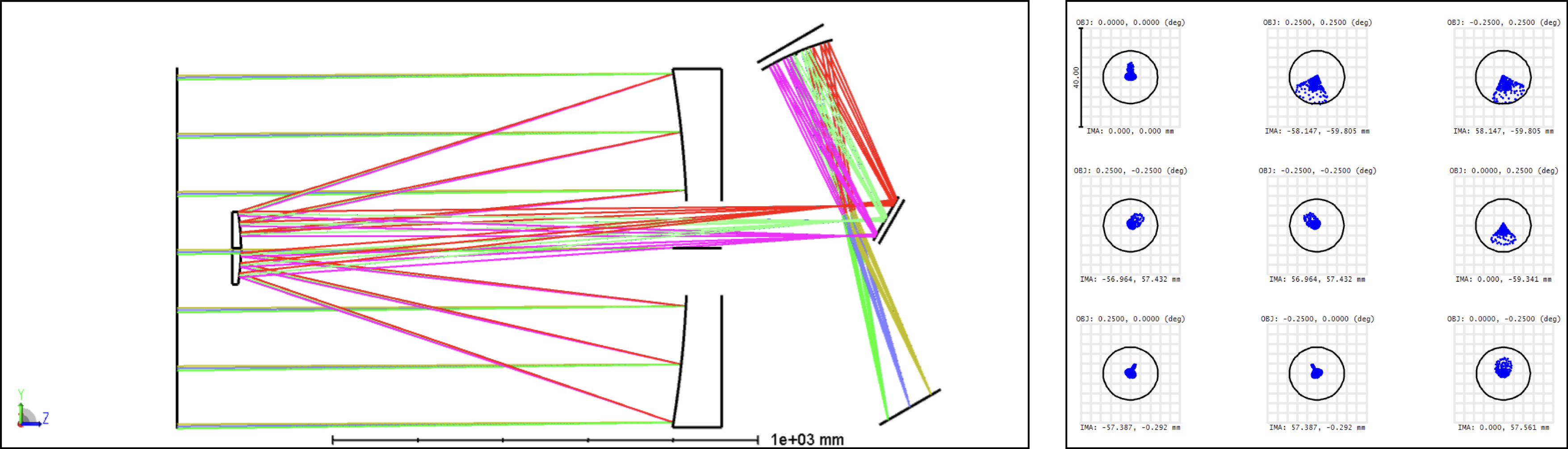}
  \caption{Optical layout of the \theia\ telescope with the spot
    diagram.}
  \label{fig:telescope}
\end{figure}

The baseline \theia Payload Module (PLM) uses the heritage knowledge
of the consortium members for space mission concepts like \gaia,
\hst/FGS, \simMission, NEAT/M3, \theia/M4+M5 and \euclid. Two
different possible concepts can be adopted. A NEAT-like mission
consisting of a single off-axis mirror sending the light onto a
detector either in a formation flight configuration\cite{Malbet+12} or
on a deployable boom, or a \euclid-like mission with a Korsch
three-mirror anastigmatic (TMA) telescope, a single focal plane and
instrument metrology subsystems\cite{Malbet+21}. Both concepts consist
in adopting a long focal length, diffraction limited, telescope and
additional metrology control of the focal plane array. The baseline
remains therefore the 0.8-m on-axis TMA telescope operating at visible
wavelengths described in the \theia/M5 proposal but with an update of
the optical design where all optics are still coaxial but with a field
of view whose center is shifted by 0.45\,deg in order for the light
beam to avoid the plane mirror after reflection on M3
(Fig.~\ref{fig:telescope}).

Compared to the proposed \theia/M5 mission concept, another progress
is the new CMOS detectors which allow up to $10^9$ small-size
($\sim 4\mu$m) pixels with well-controlled systematics,
capability of reading pre-determined windows around objects with pixel
readout at $\geq$1\,kHz rates to prevent saturation of bright
stars. Existing examples include the Sony IMX411ALR sensor with 150
Mpixels or the Pyxalis GigaPyx sensor that can reach 220 Mpixels (with
a 30k$\times$30k\,px CMOs sensor foreseen is the near future by Pyxalis
thanks to the “\emph{stitching technique}”). Such detectors would
considerably simplify the payload with a few or even a single detector
and read-out electronics instead of 24 covering the required
$\sim0.5^\circ$ field-of-view (FoV) in the focal plane array (FPA),
but also shorten the focal length to $\simeq 13\,\rm m$.

The key requirement on sub-\uas-level differential astrometric
precision for \theia's BSM implies control of all effects that impact
the relative positions of the Nyquist-sampled the
point-spread function (apparent size of $0.136''$ for a 0.8-m telescope
in the visible). The precision of relative positions determination on
the detector depends on the photon noise (limited by the 
reference stars), the geometrical stability of the focal plane array,
the optical aberrations, and the variation of the detector response
between pixels. This translates into a fundamental requirement of
$\sim 1\times10^{-5}$ pixel precision. This requirement can be relaxed
by an order of magnitude for the FSM, dominated by photon noise.

To monitor a variety of sources of distortions of the FPA, and to
allow the associated systematic errors to be corrected, \theia will
rely as baseline on metrology laser feed optical fibers placed at the
back of the nearest mirror to the detector(s).

In addition to measuring the FPA physical shape, the rest of the
telescope needs monitoring to control time-variable aberrations at a
certain level. The first estimation was at sub-$\mu$as level, but new
developments show that it may have been too conservative (see
Sect.~\ref{sec:diffastrometry}). Indeed the telescope geometry is
expected to vary, even at very stable environments such as L2 and
therefore a baseline telescope metrology subsystem was based on a
concept of linear displacement interferometers and piezo activators,
independently on each linear element of the structure, greatly
simplifying the system-level metrology in the \theia M5 proposal. In
this baseline design, telescope and FPA are cooled to $\sim130$ K and
$\sim150$ K, respectively, with a key requirement of $\sim30$ mK
stability over the integration time and for nm-level stability,
translating in a thermal control system coupling passive (V-Groove
radiation shields) with active cooling solutions (e.g. JT
coolers). However, the new method to derive the astrometry signal
using the reference stars allows us to measure the telescope
distortion described in Sect.~\ref{sec:diffastrometry} and therefore
considerably relax the initial requirement put for M5 from several hours to
fraction of a second (the frame exposure time is between 20\,ms for
the target star and 0.5\,s for the reference stars.

We outline in Sect.\ref{sec:TRL} some of the strategies that can be adopted to cope
with costs and Technology Readiness Levels (TRLs) of a few
initial mission-critical elements, while safeguarding the main science
objectives, during the assessment phase.

\subsection{Mission Profile}
\label{sec:miss-prof}
\theia needs to point at all-sky directions. L2 is the selected option
for the orbit, since it is very favorable for overall stability
simplifying thermo-elastic design issues. The \theia spacecraft will
be directly injected into a large Lissajous or Halo orbit at L2. To
avoid parasitic light from the Sun onto the telescope and the
detector, \theia spacecraft have baffles that protect them from Sun
light at angle larger than $\pm45^\circ$ from the Sun direction. The
baseline launcher is Ariane~6.2. The launch strategy would consist in
a unique burn of upper stage injecting directly the S/C onto a
L2-transfer trajectory, avoiding coasting on a parking orbit. The
observing time baseline to properly investigate the science program of
\theia is 4 years including time devoted to orbit maintenance. A total
of approximately 6 months has been estimated for the orbit transfer
including the spacecraft and instrument commissioning. From the total
of $\sim$ 35000 h dedicated for the scientific program, about 15 min
per slew will be dedicated to reconfiguration and station-keeping. Fields
containing science targets and reference stars will be observed in two
modes (FSM, BSM, see Sect. 2.7). The pointing acquisition of science
fields observed in either mode can be performed with a standard high
accuracy star tracker and the attitude measurements performed with a
fine guidance sensor (FGS).

\begin{table*}[t!]
  \centering
  \caption{TRL evaluation  and foreseen development plans for the
    \theia payload for the phase 1 proposal to the M7 call.}
  \includegraphics[width=0.99\hsize]{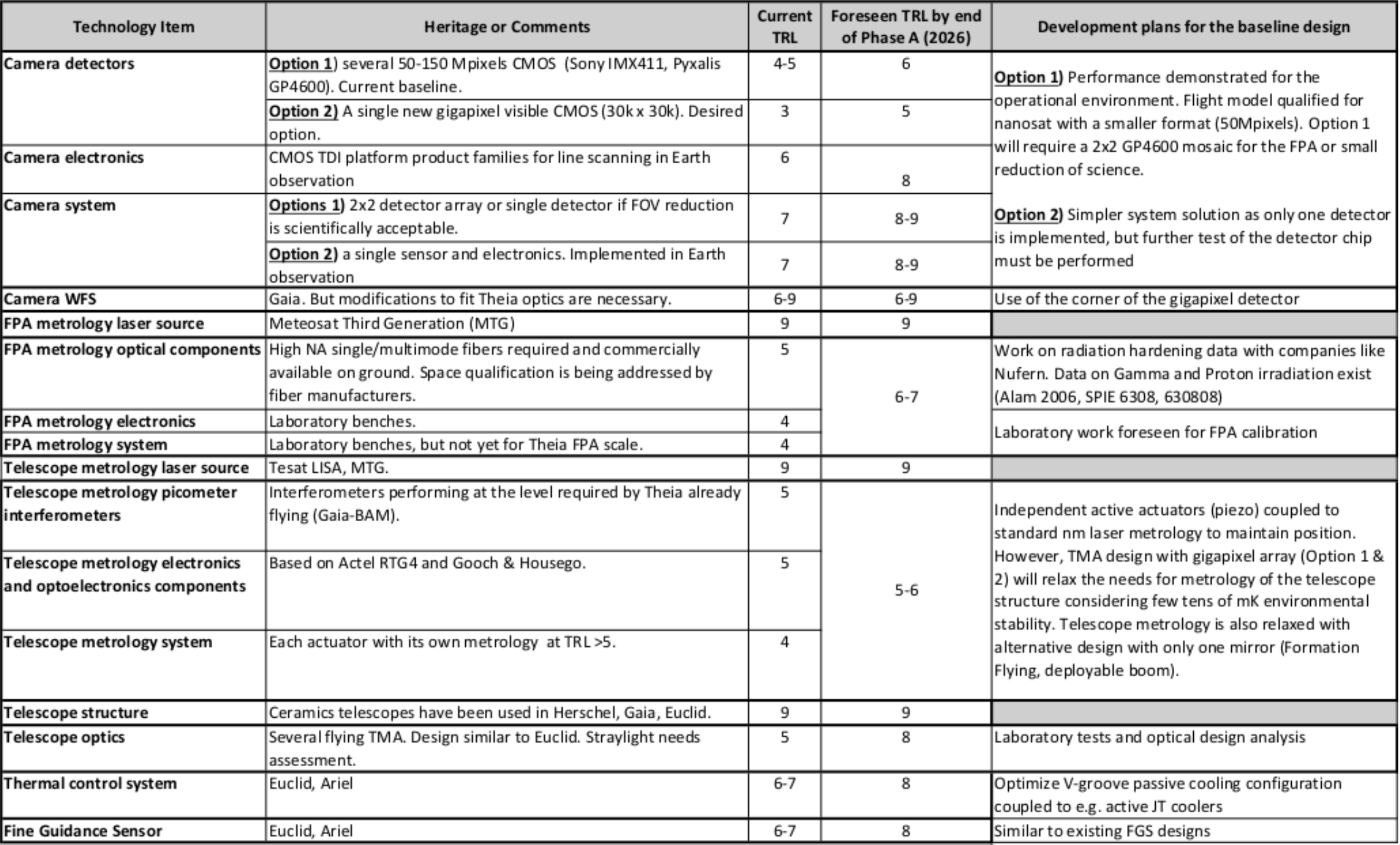}
 \label{tab:trl}
\end{table*}

\subsection{Preliminary spacecraft design}

The preliminary \theia M5 mission analyses allowed identification of a
safe and robust mission architecture mostly relying on high TRL
technologies.  The proposed mission architecture adopts the 0.8-m
Korsch TMA telescope accommodated vertically on top of a platform
including all support subsystems.

At the inital stage, a high thermal stability of the telescope was
thought to be necessary to ensure its performances. It was obtained
through the use of a Sun Shield on which is accommodated the Solar
Array and a vertical V-groove screen. The design of the satellite is
based on the \euclidx service module with a downscaled size to better
suit to specific \theia needs.  Similarly to the \euclidx and \herschelx
satellites, the \theia Korsch telescope is placed on top of the
service module in a vertical position.

\subsection{Resources and Communications}
\label{sec:resources}

The baseline M7 \theia satellite had a dry and wet mass of $\sim1000$
kg and $\sim1300$ kg, respectively, and a payload mass of $\sim400$
kg. Power resources were estimated at $\sim1400$ W for the S/C and
$\sim400$ W for the PLM. In terms of communication requirements, the
amount of science data produced by the payload module was estimated to
a total average of 95 Gbytes/day (135 Gbytes/day worst case),
including a compression factor of 2.5 with a download data rate of
$\sim$75 Mbps with an ESA 35m ground station. Consequently, daily
visibility periods of about 4 hours would be necessary, similar to
\euclid. All the above requirements fit the constraints in the
technical annex to the present Call. The foreseen simplified payload
configuration will further relax these requirements.

\subsection{TRL assessment, development plans, fall-back scenarios,
  impact on science}
\label{sec:TRL}

The current assessment of the TRL for the various key technologies
within the proposed baseline design (and identified options) is
summarised in Table~\ref{tab:trl}. The key \theia PLM profits from a
series of developments performed for past missions, but phase-A
activities will be necessary to raise the TRL of the telescope and FPA
metrology system and electronics. Phase-A activities will also be
required to raise the TRL and space-qualify large-format CMOS
detectors, to breadboard the FPA as an integrated system, and for
straylight assessment.

Phase-A activities will be needed to demonstrate the achievement of
the centroiding precision goal of $\sim1\times 10^{-5}$ pixels on CMOS
detectors. Crouzier et al.\cite{Crouzier+16} and Nemati et
al.\cite{Nemati20} have achieved 6$\times$10$^{-5}$ pixel precision in
the laboratory on a e2V CCD.  Centroiding algorithms for differential
astrometry on TESS science data have demonstrated in-flight
uncertainties of $\leq 1\times$10$^{-5}$ pixels on single frames and
$\sim 2 \times 10^{-6}$ as collective performance for an observing
sequence\cite{Gai+22}.

If the centroid goal is degraded by one order of magnitude, the core
science objectives focused on faint-source astrometry (dark matter,
cosmology, compact objects) will not be impacted, as they rely on
photon-noise limited observations. The exoplanet science case, focused
on the bright-star regime for which the ultimate systematic noise
floor achievable is critical, could accommodate up to a factor of
three in degradation of the centroiding precision goal. Relaxing the
requirement would still allow to reach sensitivity to $1.0 - 1.2$
M$_\oplus$ HZ planets around $\sim25\%$ of the nearest solar-type
stars sample (while still achieving sensitivity to 2-5 M$_\oplus$
super-Earths for the 60-target sample). This would enable in turn
detection of at least a few such companions (or at least a handful if
higher multiplicity in the HZ is considered) in the event of their
true occurrence rate matching the lower end of the estimate from
Bryson et al.\cite{Bryson+21}, but as many as $\geq10$ true Earth-like
planets (not considering higher multiplicity in the HZ) in case the
true occurrence rate exceeds the median value determined by Bryson et
al. \cite{Bryson+21}.
  
Plans for laboratory work to perform the above mentioned Phase-A
activities have been laid out. A detailed PLM design will be presented
at the time of Phase-2 proposal submission, in which the trade-off
between the baseline and the other configurations outlined here will
be resolved. A possible back-up scenario will also be presented, which
might include, e.g., an alternative configuration taking advantage of
existing, high-TRL metrology solutions, or alternative strategies for
centroiding on very bright objects (for example using diffraction
spikes).
  
\section{Management scheme}
\label{sec:management}

Over 100 researchers were originally part of the \theia M5 mission
Consortium. Presently 11 European Union countries are represented. The
core team includes members from Italy, France, Germany, Sweden, Spain,
Austria, Denmark, Portugal, and we have additional contributions from
Switzerland, The Netherlands, and Poland. Scientists from countries
outside Europe (Israel, USA) are also involved. The management team of
the \theia Consortium consists of the Co-PIs and Co-Is from each of
the main contributing countries plus the central leadership of the
Consortium: the PI is envisioned to be supported by Project Manager,
Instrument Lead, Calibration and Operation Lead, Science Ground
Segment Manager. The Co-PIs lead the national groups responsible for
the major components of the payload, Italy, France, Germany \&
Sweden. All other countries contributing either to the science case,
or to the payload development, or both, are led by a Co-I. We envision
a relevant role for the \theia Science Team, that would advise ESA on
all aspects of the mission potentially affecting its scientific
performance, assisting the ESA Project Scientist in maximising the
overall scientific return of the mission.

The \theia mission is proposed to be a fully European mission, led by
ESA. In a preliminary iteration, the \theia Consortium has converged
on a scenario for the distribution of responsibilities for providing
the following Payload Module (PLM) systems: camera + FGS, telescope
(optics, structure, thermal stabilization control), focal plane array,
all metrology subsystems, on-board control software, overall PLM
assembly, integration and verification (AIV).

Even if ESA did not select \theia\ for the  M7 call, we continue the
work to develop the key science and technical issues.

\section{Recent developments on differential astrometry measurements}
\label{sec:diffastrometry}

\begin{figure}[t]
  \centering
  \includegraphics[width=0.9\hsize]{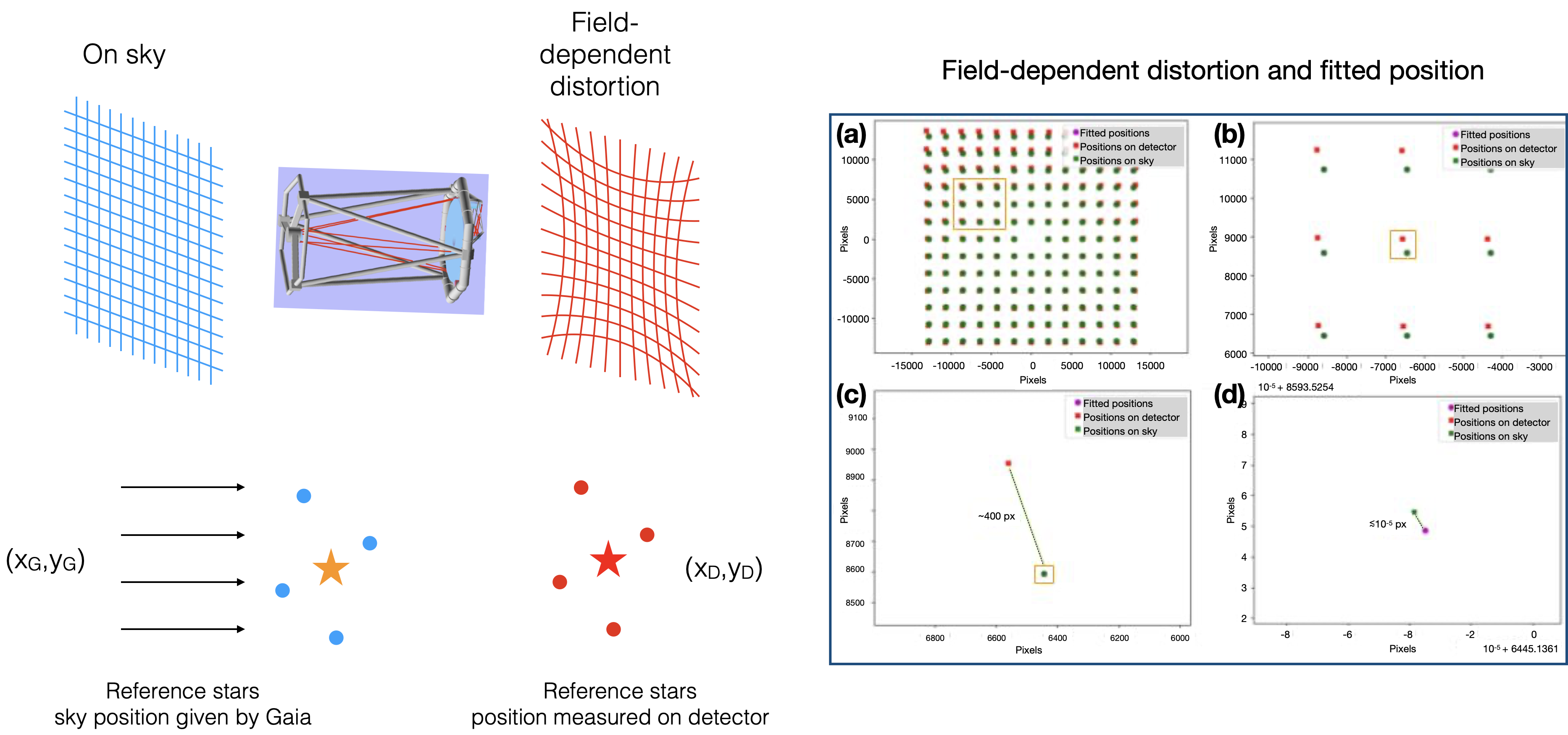}
  \caption{Calibration of the field distortion due to the telescope
    using reference stars. Left schematic of the optical set-up. The
    reference stars (blue dots) have known positions on the sky thanks
    to the Gaia catalogue. The positions of the stars measured on the
    detectors are the red dots. Thanks to the polynomial relationship
    between the positions of the reference stars on the detector and
    their positions on the sky (Gaia catalogue), it is possible to
    retrieve the sky position of the target (yellow star) from the
    detector position (red star). Right figures (a to d) show the position of
    a grid of $13\times13$ reference stars. Green is the unperturbed
    sky position, red is the detector position and magenta is the
    fitted position using the polynomial relationship. From (a) to (d)
    the 4 panels show increasing magnification.}
  \label{fig:distortion}
\end{figure}

Recently, we have undertaken a study to better understand how to reach
the accuracy of the measurements in the BSM case for exoplanets in
order to derive the requirements on the telescope design both
mechanical and optical. The principle of the measurement is summarized
on Fig.~\ref{fig:distortion}. What is observed is a target in the
middle of a 0.5\,deg\,$\times$\,0.5\,deg field where one can find many
fainter reference stars. However the telescope will move the positions
of the star with respect of an exact conjugation on the detector of
the grid on the sky, due to the different aberrations / defects of the
telescope, including the large field distortion.  What is measured is
the position on the detector $(X_D,Y_D)$ that needs to be related to
the "true" position on the sky. For the reference stars, these
positions $(X_G,Y_G)$ are known thanks to the Gaia Catalogue.

The mathematical transformation of the images of stars towards their
"true" positions on the sky can be approximated by a 2D polynomial of
order $n$.  Explicitly, if $X$ and $Y$ correspond to an approximation
of the "true" coordinates, one reads :
\begin{equation}
  \label{eq:1}
  \left\{
    \begin{array}{llllll}
        X &= &P\left(\mathbf{A},X_{D},Y_{D}\right) &= &\sum\limits_{\underset{i+j\le n}{i,j=0}}^{n} A_{ij}X_{D}^i\,Y_{D}^j \\
        Y &= &P\left(\mathbf{B},X_{D},Y_{D}\right) &= &\sum\limits_{\underset{i+j\le n}{i,j=0}}^{n} B_{ij}X_{D}^i\,Y_{D}^j 
    \end{array}
\right.
\end{equation}
with $\mathbf{A}$ and $\mathbf{B}$ the coefficients of this
polynomial. There are therefore a total of $(n+1)(n+2)$ coefficients
to determine. Using a classical Levenberg-Marquardt
method\cite{Levenberg+44,Marquardt+63}, the
position of the reference stars can be fitted to their known $(X_G,Y_G)$ positions
on the sky and therefore determine the values of these coefficients
$\mathbf{A}$ and $\mathbf{B}$. The right part of
Fig.~\ref{fig:distortion} shows in green color a grid of
$13\times13$ stars around the telescope field center and in red color
the barycenters of the spot diagrams of Zemax ray tracing on the
detector. The positions in magenta are the result of the fitting of a
polynomial with order $n=7$ and 72 unknowns.  In panel (c), the
distance between the detector position and the true position can be of
the order of several hundreds of pixels, whereas in panel (d) the
distance of the fitted and true position is only a small fraction of a
pixel ($\simeq10^{-5}$\,pixel).

\begin{figure}[b]
  \centering
  \includegraphics[width=0.9\hsize]{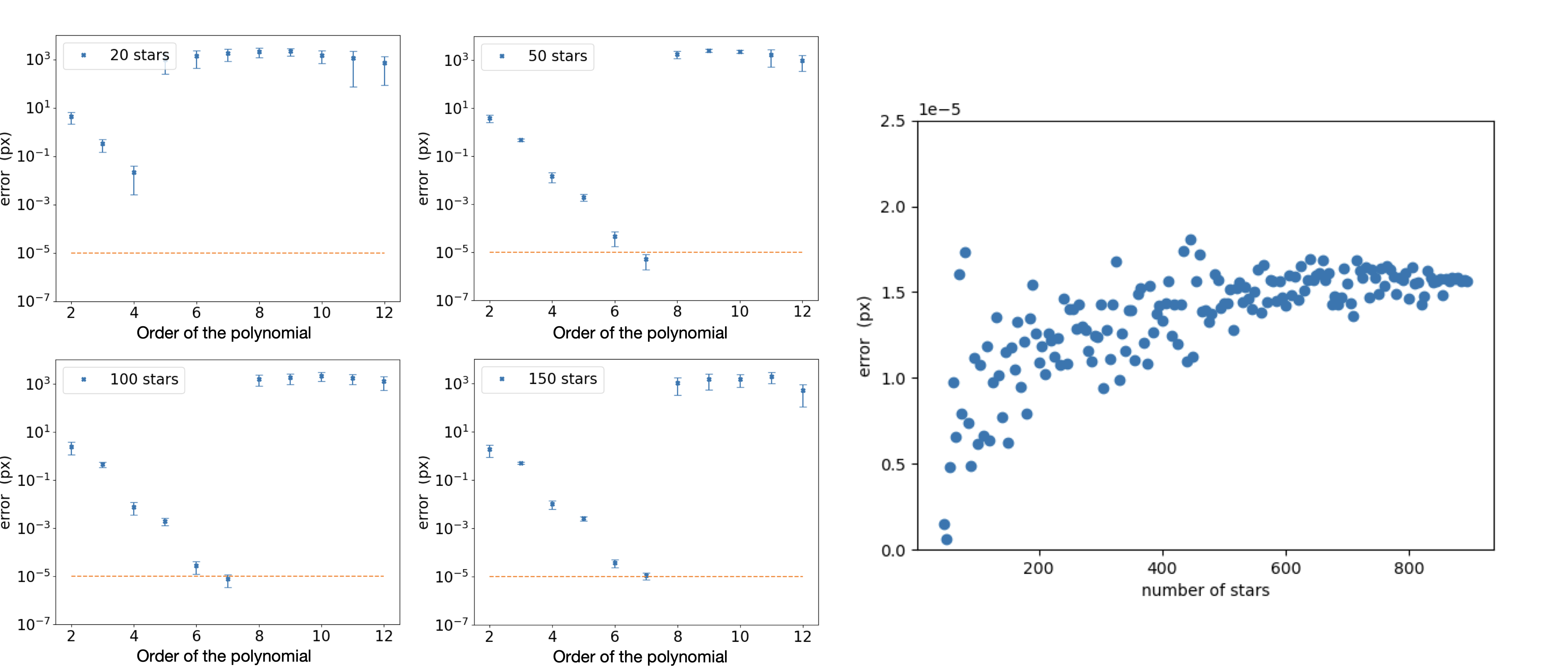}
  \caption{Measurement errors on the target position at the center of
    th field in function of the order of the polynomials and number of
    reference stars (left panels).  The right window shows how this
    accuracy changes with the number of reference stars used with a 2D
    polynomial of order $n=7$.}
  \label{fig:target-fitting-accuracy}
\end{figure}

Using a linear combination of the reference stellar positions, e.g.\
their barycenter with adequate weighting, a reference point can be
located close to the target image.  With the polynomial fit, we can
then estimate the sky position of the target, and make the astrometric
measurement between the target and this reference point.  An important
issue is the accuracy of the
fit. Figure~\ref{fig:target-fitting-accuracy} shows that the error on
the position decreases exponentially with $n$ and goes ultimately
below $10^{-5}$\,pixels for $n=7$. This accuracy does not change
significantly with the number of reference stars if the number of
constrains $N_\mathrm{stars}$ is greater than half the corresponding
number of parameters, $(n+1)(n+2)/2$, as can be seen from the right
panel of Fig.~\ref{fig:target-fitting-accuracy}.

We have also re-computed the ray tracing and the fitting procedure
with an optical configuration of the telescope where the M2 mirror has
been tilted by one arcsecond. The distorted positions are found far
away from the initial value (several hundreds of picels), but the
estimated position in the field center neighborhood remains determined
at the $10^{-5}$\,pixel pixel precision, pointing out the capability of the
fitting procedure to correct for significant instrumental defects.

However, we need to make the 2D polynomial computation more robust at
higher orders than $n=7$. Indeed, with $13\times13\,-\,1=168$
reference stars, a polynomial with a maximum of $336$\,parameters
should be determined corresponding to a polynomial of order of $n\leq16$
unless there are redundancies not yet understood.

We plan now to repeat our analysis using Fourier series as well as
Zernike polynomials.  A last question still to be investigated is the
best strategy to define a reference barycenter close to the center of
the field where the target star is located and with its actual field
of view. Our plan is to use realistic reference stars and to take into
account both their positions in the field and the photon noise of
their image. We are also going to investigate crowded fields as the cores
of globular clusters.

We conclude that the fitting procedure using polynomials can correctly
take into account for the optical distortion of the nominal TMA
telescope. This can also compensate systematic astrometric errors
caused by a 1\,arcsec tilt to the M2 mirror (corresponding to
$1\,\mu$m displacement of one side of the mirror structure), showing
that there is no need for a very precise metrology control of the
telescope.  In other words, the type of metrology that we use for the
telescope structure is not based on metrology lasers located on the
mechanical structure itself but is funded on the position on the
detector of reference stars compared to their actual position on the
sky as determined by Gaia. Our measurements are buit on what we can
call a \emph{reference star metrology}.

\section{Conclusion}

Theia is a project for an astrometric observatory based on high-precision
differential astrometry measurements on a limited single field
(0.5\,deg$\times$0.5\,deg). The two main science cases are nature of
dark matter and Earth-like planets around solar neighbourhood. Theia’s
payload is a 0.8\,m diameter diffraction limited TMA Korsch
telescope. New very large format CMOS visible detectors are being
investigated in order to cope with a diffraction-limited yet
large field of view without too many detectors. Telescope stability impacts
performances except if one is able to calibrate the telescope
distortion using reference stars with Gaia positions. Initial results
show that with 50 to 100 reference stars whose position is known with
Gaia, one can calibrate the center of the field to an accuracy lower than
the $10^{-5}$ pixel .   

\appendix

\bibliography{theia-spie-2022} 

\begin{thebibliography}{10}

\bibitem{Malbet+12} {Malbet}, F., {L{\'e}ger}, A., {Shao}, M., et al.,
  ``High precision astrometry mission for the detection and
  characterization of nearby habitable planetary systems with the
  nearby earth astrometric telescope (neat),'' {\em Exp. Ast.}~{\bf
    34}, 385--413 (Oct. 2012).

\bibitem{Malbet+21} {Malbet}, F., {Boehm}, C., {Krone-Martins}, A., et
  al., ``Faint objects in motion: the new frontier of high precision
  astrometry,'' {\em Exp. Ast.}~{\bf 51}, 845--886 (June 2021).

\bibitem{Moore+99}
{Moore}, B., {Ghigna}, S., {Governato}, F., {Lake}, G., {Quinn}, T., {Stadel},
  J., and {Tozzi}, P., ``Dark matter substructure within galactic halos,'' {\em
  \apjl}~{\bf 524},  L19--L22 (Oct. 1999).

\bibitem{Feldmann&Spolyar15}
{Feldmann}, R. and {Spolyar}, D., ``Detecting dark matter substructures around
  the milky way with gaia,'' {\em \mnras}~{\bf 446},  1000--1012 (Jan. 2015).

\bibitem{Navarro+96}
{Navarro}, J.~F., {Frenk}, C.~S., and {White}, S. D.~M., ``The structure of
  cold dark matter halos,'' {\em \apj}~{\bf 462},  563 (May 1996).

\bibitem{Pontzen&Governato12}
{Pontzen}, A. and {Governato}, F., ``How supernova feedback turns dark matter
  cusps into cores,'' {\em \mnras}~{\bf 421},  3464--3471 (Apr. 2012).

\bibitem{Read+21}
{Read}, J.~I., {Mamon}, G.~A., {Vasiliev}, E., {Watkins}, L.~L., {Walker},
  M.~G., {Pe{\~n}arrubia}, J., {Wilkinson}, M., {Dehnen}, W., and {Das}, P.,
  ``Breaking beta: a comparison of mass modelling methods for spherical
  systems,'' {\em \mnras}~{\bf 501},  978--993 (Feb. 2021).

\bibitem{Jing&Suto02}
{Jing}, Y.~P. and {Suto}, Y., ``Triaxial modeling of halo density profiles with
  high-resolution n-body simulations,'' {\em \apj}~{\bf 574},  538--553 (Aug.
  2002).

\bibitem{Hills88}
{Hills}, J.~G., ``Hyper-velocity and tidal stars from binaries disrupted by a
  massive galactic black hole,'' {\em \nat}~{\bf 331},  687--689 (Feb. 1988).

\bibitem{Hawking71}
{Hawking}, S., ``Gravitationally collapsed objects of very low mass,'' {\em
  \mnras}~{\bf 152},  75 (Jan. 1971).

\bibitem{Ricotti&Gould09}
{Ricotti}, M. and {Gould}, A., ``A new probe of dark matter and high-energy
  universe using microlensing,'' {\em \apj}~{\bf 707},  979--987 (Dec. 2009).

\bibitem{Frampton+10}
{Frampton}, P.~H., {Kawasaki}, M., {Takahashi}, F., and {Yanagida}, T.~T.,
  ``Primordial black holes as all dark matter,'' {\em \jcap}~{\bf 2010},  023
  (Apr. 2010).

\bibitem{Bringmann+12}
{Bringmann}, T., {Scott}, P., and {Akrami}, Y., ``Improved constraints on the
  primordial power spectrum at small scales from ultracompact minihalos,'' {\em
  \prd}~{\bf 85},  125027 (June 2012).

\bibitem{Moore+17}
{Moore}, C.~J., {Mihaylov}, D.~P., {Lasenby}, A., and {Gilmore}, G.,
  ``Astrometric search method for individually resolvable gravitational wave
  sources with gaia,'' {\em \prl}~{\bf 119},  261102 (Dec. 2017).

\bibitem{GarciaBellido+21}
{Garcia-Bellido}, J., {Murayama}, H., and {White}, G., ``Exploring the early
  universe with gaia and theia,'' {\em \jcap}~{\bf 2021},  023 (Dec. 2021).

\bibitem{Vitral+22}
{Vitral}, E., {Kremer}, K., {Libralato}, M., {Mamon}, G.~A., and {Bellini}, A.,
  ``{Stellar graveyards: clustering of compact objects in globular clusters NGC
  3201 and NGC 6397},'' {\em \mnras}~{\bf 514},  806--825 (July 2022).

\bibitem{Meunier&Lagrange22}
{Meunier}, N. and {Lagrange}, A.~M., ``{A new estimation of astrometric
  exoplanet detection limits in the habitable zone around nearby stars},'' {\em
  \aap}~{\bf 659},  A104 (Mar. 2022).

\bibitem{Bryson+21} {Bryson}, S., {Kunimoto}, M., {Kopparapu}, R.~K.,
  et al., ``The occurrence of rocky habitable-zone planets around
  solar-like stars from kepler data,'' {\em \aj}~{\bf 161}, 36
  (Jan. 2021).

\bibitem{Lindegren+2018}
{Lindegren}, L., {Hern{\'a}ndez}, J., {Bombrun}, A., et al., ``{Gaia Data Release 2. The astrometric solution},'' {\em
  \aap}~{\bf 616},  A2 (Aug. 2018).

\bibitem{sanderson+2019} {WFIRST Astrometry Working Group},
  {Sanderson}, R.~E., {Bellini}, A., {Casertano}, S., et al.,
  ``{Astrometry with the Wide-Field Infrared Space Telescope},'' {\em
    J. of Astr. Teles., Instr., and Syst.}~{\bf 5}, 044005
  (Oct. 2019).

\bibitem{Gravity+2021} {Gravity Collaboration}, {Abuter}, R.,
  {Amorim}, A., {Baub{\"o}ck}, M., et al., ``{Improved GRAVITY
    astrometric accuracy from modeling optical aberrations},'' {\em
    \aap}~{\bf 647}, A59 (Mar. 2021).

\bibitem{Crouzier+16}
{Crouzier}, A., {Malbet}, F., {Henault}, F., et al., ``A detector interferometric calibration experiment for high
  precision astrometry,'' {\em \aap}~{\bf 595},  A108 (Nov. 2016).

\bibitem{Nemati20}
{Nemati}, B., ``Photon counting and precision photometry for the roman space
  telescope coronagraph,'' in [{\em SPIE Conf.
  Series}{\nolinebreak\hspace{0.1em}]},   {\bf 11443} (Dec. 2021).

\bibitem{Gai+22} {Gai}, M., {Vecchiato}, A., {Riva}, A., {Busonero},
  D., {Lattanzi}, M., {Bucciarelli}, B., {Crosta}, M., and {Qi}, Z.,
  ``{Astrometric Precision Tests on TESS Data},'' {\em \pasp}~{\bf
    134}, 035004 (Mar. 2022).

\bibitem{Levenberg+44}
Levenberg, K., ``A method for the solution of certain non-linear problems in
  least squares,'' {\em Q. Appl. Math.}~{\bf 2},  164--168 (1944).

\bibitem{Marquardt+63}
Marquardt, D.~W., ``An algorithm for least-squares estimation of nonlinear
  parameters,'' {\em Journal of the Society for Industrial and Applied
  Mathematics}~{\bf 11}(2),  431--441 (1963).

\end{thebibliography}
\bibliographystyle{spiebib} 

\end{document}